# On-Demand Networking for Ubiquitous Connectivity and Network Resilience: A Network-in-a-Box Solution

Ki-Hong Park, *Senior Member*, IEEE, Mohamed-Slim Alouini, *Fellow*, IEEE and Yunfei Chen, *Senior Member*, IEEE

*Abstract*— Recently, the wireless community has initiated research on the sixth generation (6G) cellular network for the next decade. The 6G visions are still under development but are converging toward ubiquitous, sustainable, and automated digital society. A network-in-a-box (NIB) is a portable and fully-fledged networking solution that has many potentials to stimulate 6G visions, especially for ubiquitous and resilient network connectivity. In this article, we highlight how NIB features suit 6G use cases and requirements and how it can be used for 6G communications. In addition, we discuss the challenges of the potential enabling technologies of 6G that can reinforce the NIB performance.

## I. Introduction

ON 25 September 2015, 193 world leaders gathered at the United Nations (UN) promise to the sustainable development goals (SDG) that end societal, economical, and environmental injustice and inequality by 2030 [1]. In order to surmount the social injustice and economical inequality for UN SDGs in low income area in cities, remote areas and developing countries, information and communication technology (ICT) will help fill the economical divide and social disparity through the provisioning of technological breakthrough, e.g., extended reality (XR), e-health, remote education, etc. In this regard, the sixth generation (6G) of broadband cellular networks has been anticipated to become an important tool to cope with the challenges facing by UN SDGs [2].

The driving 6G technologies currently being proposed are shaped by the use cases embracing the service-specific demands from a wide range of industry verticals and communities. Generally speaking, they will bond together for ultra-massive and ubiquitous network connectivity of everything, i.e., people and any type of machines, anytime and anywhere in compliance with their demands to provide communication, control, computing, localization and sensing (3CLS) services [3]. To do so, ultra-high mobile broadband service is required to exchange a massive amount of data among end users under control. Virtualization of network functions efficiently orchestrate and customize network resources to meet the service-centric traffic demands. With the substantial increase of data and the proliferation of smart devices in 6G, current quantum leap of artificial intelligence (AI) and machine learning (ML) techniques will contribute to the network self-sustainability and autonomy through diverse network layers [4].

To support the ubiquitous network connectivity, 6G network has to be resilient to not only the conditions that do not fulfill the acceptable level of key performance indicators (KPIs), e.g., reliability and latency in the network, but also the circumstances externally confronting the defect and challenges to the ordinary operation of networks. For instance, the network should be recovered quickly for rescue missions and the population in a damaged area where the network infrastructure is partially destroyed by natural disasters. In some temporary events with excessive population, e.g., concerts and sport events, mobile network operators (MNOs) can experience traffic congestion and signaling storm in some access points and should temporarily deploy extra access points nearby. As such, network resilience in 6G should include the coverage areas where the environments is unpredictable and for this the network deployment needs to be flexible and adaptable.

On the other hand, in recent years, the industry and academia have dedicated to developing a single or handful of portable equipment consisting of all software and hardware blocks required to configure or reconfigure a mobile network, which is called a network-in-a-box (NIB) [5]. An NIB is easy to deploy to form a mobile network. It is specially advantageous for emergency and tactical networks requiring fast deployment and configuration. An NIB can either work alone or is interoperable with other NIBs and incumbent network components to provision end-user services as well as backhauling through multiple air and wired interfaces. In addition, the industry strives to embody a high level of autonomy in NIB, i.e., self-organization mainly featuring three capability, self-configuration, self-optimization, and self-healing. With this autonomy, NIB can facilitate to implement the network

K.-H. Park, and M.-S. Alouini are with the Computer, Electrical, Mathematical Sciences and Engineering Division, King Abdullah University of Science and Technology (KAUST), Thuwal, Makkah Province, Kingdom of Saudi Arabia (Email: kihong.park@kaust.edu.sa; slim.alouini@kaust.edu.sa).
Y. Chen is with the School of Engineering, University of Warwick, Coventry, CV4 7AL, United Kingdom (Email: yunfei.chen@warwick.ac.uk).

flexibility and to expand the ubiquitous connectivity for 6G mobile networks.

In this article, we discuss how NIB fits 6G mobile networks in terms of requirements, services, scenarios and challenges. We will briefly review the 6G visions, use cases, and its KPI requirements which are currently under discussion. Then, we will analyze how the NIB features assist the 6G vision and implement 6G use cases in practice by tackling the 6G challenges. We will also discuss how NIB can embody 6G technologies. The discussion will be validated by a use case of NIB on maritime communication network as an example. Consequently, the main contribution of this magazine paper is two-fold, i) shedding light on a concrete and forward-looking vision of NIB in 6G mobile network and ii) sharing a timely research milestone, challenges, and recommendations to expedite the leap from traditional NIB toward 6G-NIB.

The rest of this article is organized as follows. Section II briefly reviews the 6G visions, use cases, requirements, and potential driving technologies. In Section III, we introduce the concept of NIB and its technological features, while showing how NIB suits the 6G visions. We present in Section IV the case study on spectral efficiency enhancement under technological flexibility in NIB-based maritime network. The challenges for realizing 6G driving NIB will be given in Section V and, finally, some concluding remarks are drawn in Section IV.

## II. 6G VISIONS, REQUIREMENTS, AND TECHNOLOGIES

In this section, we will briefly review the 6G initiatives in terms of its visions, use cases, requirements, and technologies. The details on 6G initiatives described here will be focusing on aspects related to the NIB. For the readers interested in more information on the general conceptual study on 6G, please refer to [2]-[4], [6]-[8].

### A. 6G Visions

The visions of 6G mobile network are mainly two-fold, i) overcoming the limitations of 5G mobile networks, and ii) providing ubiquitous, autonomous, and intelligent connectivity between network entities. With the proliferation of Internet of Things (IoT) devices and smart devices and the increase of data-intensive use cases, it is foreseen that 5G mobile networks cannot bear the unstoppable growth in data traffic near 2030. To deal with this, the spectrum in 6G is expected to move to a higher band at millimeter wave (mmWave) and terahertz (THz) frequency to support data rates hundreds or thousands times larger than 5G. On another front, 6G will realize the global connectivity of everything for access to the internet. The components in 6G mobile networks will be evolved to autonomously and intelligently manage the service-specific contextual requirements and sustain the network efficiency on spectrum, energy, and computing. 6G will become a multi-functional system that jointly integrates multiple functions with respect to 3CLS services to ideally realize the zero-instant and zero-touch decision making of multi-purpose applications in network [8].

### B. Potential 6G Use Cases and Requirements

6G study groups are expecting that 5G cannot bear to support the unprecedented requirements of new use cases in the near future. Moreover, enhanced version of traditional 5G applications will still keep their importance in 6G. We list some examples of potential applications that will call for the capability beyond 5G as follows:
- Extreme capacity backhaul and fronthaul
- Multi-sensory XR and holographic telepresence
- Autonomous industrialization and mobility
- Body area networks using wearable devices and implants
- Global connectivity
- Blockchain and distributed ledger technology.

The new 6G use cases will bring big hurdles to the current mobile networks and among them, the biggest challenge will be to serve a massive amount of data traffic in the network which will be expected to promote a peak data rate up to hundreds times above the 5G requirements. As industrialization become more and more automated, connected robotics and mobility will require ultra-low latency and extremely low reliability for real-time operation and safety. Multi-sensory XR and smart body implants might introduce new momentous barometers of network performance, e.g., latency jitter and perceptual requirements. The speculative KPIs currently estimated by 6G initiative groups are listed as follows.
- Peak data rate: 1 Tb/s
- Experienced data rate: 1 Gb/s
- Peak spectral efficiency: 60 b/s/Hz
- User spectral efficiency: 3 b/s/Hz
- Maximum bandwidth: 100 GHz
- Operating frequency: up to 1 THz
- Area traffic capacity: 1 Gb/s/m$^2$
- Connection density: $10^7$ devices/km$^2$
- Reliability: 99.99999%
- Latency: 10 µs
- Mobility: 1000 km/h
- Energy efficiency: 1 Tb/J
- Jitter: 1 µs
- Localization precision: 1 cm on 3D

The 6G KPIs mentioned above are the values that 6G initiatives are currently expected to realize the new 6G use cases. The values might be slightly different across different 6G initiatives. Some 6G initiatives suggest that the area traffic capacity and connection density in 6G should be considered in three-dimensional space to integrate aerial vehicles into a mobile network. Moreover, some KPIs such as energy efficiency, jitter, and localization accuracy are initially taken into consideration as major factors in 6G which are required to realize new applications, e.g., control-driven applications, multi-sensory systems, autonomous applications, etc.

### C. Potential 6G Driving Technologies

The evolved performance indicators for 6G will lead to the emergence and convergence of new driving technologies to fit the network on service-oriented demands. The 6G network will presumably feature the following trends to speed up the killer technologies involved.

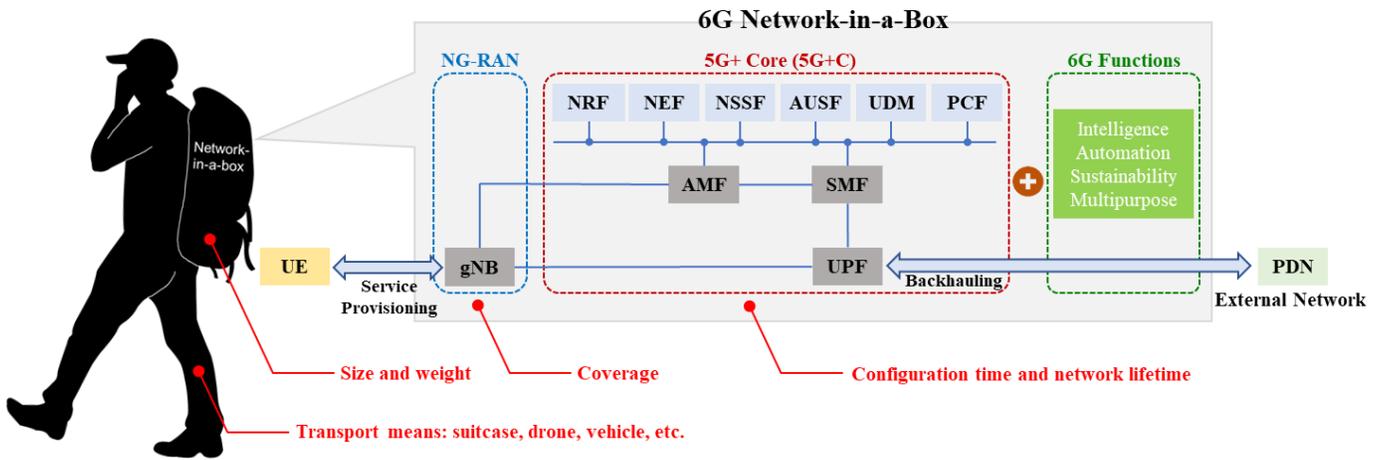

*Figure 1. The conceptual structure and the key factor of an easy-deployable NIB potentially applicable to 5G and beyond (5G+) that make an NIB different from traditional wireless network. The acronyms of network functions in 5G can be found in [9], while UE, gNB and PDN stand for user equipment, new generation Node B, and packet data network. This structure also works for future 6G.*

1) **6G will provide ubiquitous and seamless connectivity in globe [2]**

6G will be the first generation to cover the digital divide over the world. Satellite and non-terrestrial communications will be integrated to provide broader coverage in a three-dimensional space. Millions of smart devices will connect to a network, and novel multiple access, e.g., scalable cell-free access, will be adopted to manage dense connection more efficiently.

2) **6G will operate at a higher frequency to achieve larger bandwidth [6]**

The massive amount of data traffic will bring in new radio frequency (RF) spectrum bands allocated at mmWave and THz frequency range, which requires the development of new transceiver architectures. Furthermore, optical wireless communication (OWC) working on visible light and infrared wavelength can be complementary to the traditional RF communications.

3) **6G will be intelligent [7]**

The innovation in advanced technologies such as autonomous driving/flying vehicles, industry 4.0, XR, etc. will succeed when being anchored up with the advances in AI and ML techniques. The availability, computation, and exchange of massive dataset in a network will require the intelligence in 6G. The ML will play an important role to solve such new challenges through the 6G network from physical layer to application layer. The massive dataset for computation should be intelligently distributed into edge cloud or central cloud in a network.

4) **6G will tackle the convergence of multi-purpose services [8]**

In order to jointly satisfy more stringent requirements regarding 3CLS-oriented services, 6G will converge to a more general multi-purpose network.

5) **6G will be secure and resilient [3]**

The unprecedented availability of massive data in 6G applications might lead to vulnerability in security. The self-driving vehicles and autonomous systems require extra-security for safety when operating in a network. The ML-based security solution might be indispensable. For ubiquitous connectivity, 6G should be flexible enough to sustain the network for the cases of unexpected network failure such as natural disaster and on-demand services such as tactical network and data offloading due to data traffic storm.

The NIB is the mobile all-in-one network solution which integrate a core network, base station units, such as remote radio head and baseband units, into a kind of portable box or backpack. It can be a suitable solution that maintains the network connectivity in the aforementioned situations, e.g., after-disaster scenario and service provisioning in harsh environments.

Therefore, NIB is an inevitable pieces toward the ubiquitous global connectivity and network resilience in 6G. In the next section, we will introduce the concept of NIB and explain how the NIB features can suit the visions in 6G as well as how 6G driving technologies can boost the NIB features in 6G.

III. NETWORK-IN-A-BOX FOR 6G

NIB is conceptually designed as a single portable device containing all software and hardware modules to provide wireless connectivity for a group of end users, as shown in Fig. 1, and it is evolved to configure mobile ad-hoc networks (MANETs) with other NIBs. The multiple devices connected in an NIB can communicate with each other, while accessing the internet through a core network in the NIB. Transportability of NIB enables the network to be deployed and moved anywhere and anytime in an on-demand fashion, which is also known as its catchphrase "build your own coverage". Therefore, an NIB should be lightweight to transport and configure-free to deploy. In some scenarios, such as natural disaster and private/tactical networks, an NIB should have a user-friendly interface for non-expert MNOs to deploy and maintain, which means that it is a "do-it-yourself" solution.

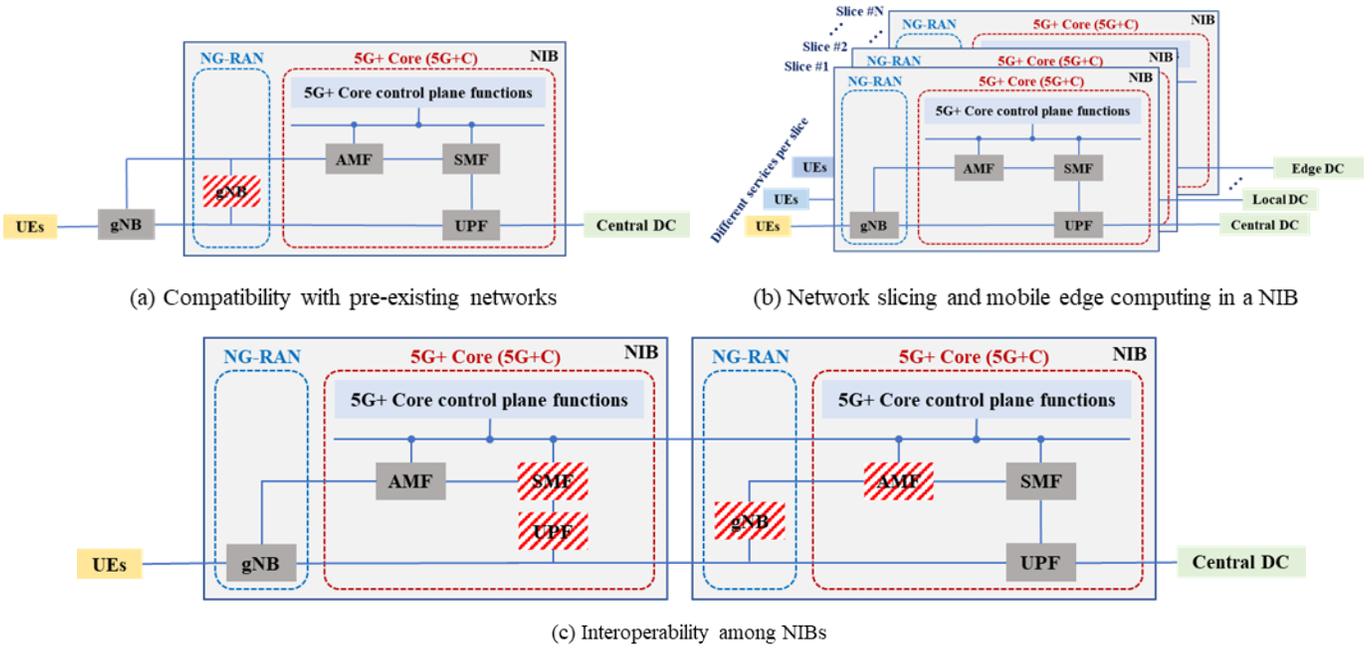

Figure 2. Network flexibility stimulated by NIBs in different scenarios. (a) In after-disaster scenario, NIB can replace the damaged network components with part of it. (b) The different services with specific demands can be efficiently provisioned via network slicing and mobile edge computing in a NIB. (c) NIBs can cooperate with each other to balance the load between them.

*A. Use Cases*

NIB is perfectly fit-for-purpose to build a sporadic on-demand network due to its physical features on size and weight. More specifically, the NIBs are suitable for the following representative applications;

1) **Emergency network**

When a natural disaster destroys the mobile network infrastructures, wireless connectivity in the damaged area should be restored fast for both rescue team and survivors for safety concerns. NIB can be deployed with different transport means within an acceptable network configuration time after disaster.

2) **Connectivity in challenging circumstances**

The mobile network is often deficient in harsh environments, such as mountainous, desertic, and maritime areas, where it is troublesome to deploy fixed network infrastructures. The NIB can resolve the data demand in these challenging areas due to its portability and cost efficiency.

3) **Rural connectivity**

The remote area is often considered profitless for MNOs due to low avenue per user and is significantly constrained on internet access. The NIB with limited and user-friendly hardware and software capability can significantly save capital and operation expenditure in a mobile network which is advantageous for local operators in rural areas.

4) **Tactical network**

A military mission and expedition in polar regions require the wireless connectivity within operation areas or to the internet, while hardly utilizing public network infrastructures. The NIB can conveniently address the problems to build tactical networks since it is easy for deployment and transportable.

5) **Private network**

For some enterprises, e.g., mining, logging, and oil companies, the work places might be changed regularly and the company-specific services in a private enterprise network are the main factors in network data traffic load. NIB offers portability and cost-efficiency to maintain the private network by minimizing the need on the public network infrastructures.

6) **Private network**

The NIB can be an attractive ready-to-use option to prepare for the intermittent increase of data traffic exceeding the network peak capacity. During the events with massive popularity, such as sport events, concerts, and festivals, NIB can be leveraged to offload the traffic as a one-off solution.

*B. Technologies in NIB*

As shown in Fig. 2, NIB communicates with end users to provide service-specific wireless connectivity, while backhauling the data traffic to the external packet data network (PDN), e.g., internet. Moreover, NIB can communicate with other NIBs to control the load balance in the network or gear into the existing public network infrastructures to fulfill the defected network functionality, as shown in Fig. 2a. Thus, depending on the deficiency of network functions, NIB can flexibly play an important role in the network. In the following, we describe in more details the technologies used in NIB from a network-functional perspective.

1) **Service provisioning**

An NIB is equipped with one or more radio interfaces to provide wireless connectivity for end users. For the after-disaster scenario, the dedicated radio technologies for public safety, such as P25, TETRA, or TETRAPOL, were leveraged in NIBs. Generally, current cellular 5G network and its backward compatible networks, e.g., 2G/3G/4G are adopted for connectivity provisioning, while Wi-Fi serves data traffic around relatively short range in an NIB-embodied network. Many types of NIBs support Wi-Fi and one or more cellular technologies simultaneously in order to maximize connectivity in the network.

2) **Backhauling**

The end users are connected to an external network through an NIB core network. Sometimes, an NIB is wired by a cable to link the pre-existing network infrastructure. In many cases, satellite communication technology is utilized for backhauling thanks to the ubiquitous connectivity over challenging circumstances where fixed network infrastructures are missing, e.g., remote area and maritime network. Other backhauling options, such as 2G-5G cellular technologies, Wi-Fi, microwave, and potentially mmWave and higher frequency communication, can be leveraged in an NIB at the same time.

3) **Interoperability**

The NIB can flexibly play different roles in a network. As mentioned above, the NIB can be working standalone to build a fully-fledged network on its own. On the other hand, it can also cooperate with other network entities, such as other NIBs or incumbent systems, to independently provide the functions of NIB in a network or to take over a part of the network functions. Multiple NIBs can communicate with each other through standard interfaces, and dynamically adjust the load balancing on 3CLS, and assign the roles in a network, as shown in Fig. 2c.

4) **Network flexibility**

As mentioned above, we highlight that the NIB-based network can adaptively support high level of network flexibility over multiple radio technologies for service provisioning and backhauling which highly requires the softwarization of radio interfaces through programmable array, general purpose processors as well as reconfigurable antennas. Ultimately, an NIB endeavors to operate without relying on a specific radio technology to sustain the network flexibility. Therefore, by implementing network function virtualization (NFV), it allows the heterogeneity in a network and enables the support of service-specific data requirement on individual network slice, as shown in Fig. 2b.

5) **Edge computing**

As the traffic load in the network increases, the data should be logically routed according to the services depending on the application verticals. NIB can directly deal with the local services or route them to local cloud server instead of accessing the central cloud server in the external network. By doing so, NIB avoids unnecessarily sending the traffic from/to a core network and reduces the latency by localizing the computational load around the end users.

6) **Self-sustainability**

An NIB aims to exhibit self-sustainable network (SSN) feature since most use cases of NIB-based networks require ease of operation by non-professionals, such as rescue team and explorers, in the deficient environments where the pre-existing network infrastructures are damaged or missing. Therefore, an NIB requires high level of autonomy on network configuration, optimization, and healing. For example, the NIBs can automatically select a main NIB node that plays a core network, adaptively reallocate the resources for load balancing among NIBs, dynamically reassign the different functions on NIBs for new NIB activation or network component failure [10].

TABLE I: System parameters for simulations

|  | Satellite | Cellular (3G, 4G) | Shipborne WiFi |
|---|---|---|---|
| Tx power | 49 dBm | 43 dBm | 20 dBm |
| Tx antenna gain | 52 dBi | 15 dBi | 10 dBi |
| Rx antenna gain | 30 dBi | 0 dBi | 10 dBi |
| Channel model | FSPL [11][1] | 3GPP [12][2] | ITU-R [13][3] |
| Path loss exponent | 2 | 3.4 | 2 |
| Carrier frequency | 20 GHz | 2 GHz | 2.4 GHz |
| Bandwidth | 5 MHz | 5 MHz | 20 MHz |

[1] Geostationary satellite free space path loss (FSPL) model.
[2] 3GPP macro cell propagation model for rural area.
[3] ITU-R line-of-sight model for IEEE 802.11n.

*C. How NIB envisions 6G?*

In the previous sections, we have specified the features of NIB and 6G from a technological point of view. Next, we will question for ourselves on how NIB can encourage and redound to the visions of 6G. We will try to seek the answer from the following concrete and inevitable research trends pursued by NIB which is related to the 6G visions.

- The NIB-based network-on-demand aims to solve the global connectivity problem.
- The physical features of NIB are suitable to provide ubiquitous connectivity in harsh environments.
- An NIB is mobile to deploy anywhere to dynamically optimize the network configuration.
- Network resilience can be strengthen by proper deployment of NIBs in an on-demand fashion.
- The technological features in NIBs are advantageous to apply to implement an integrated space-air-terrestrial network.
- Wireless backhauling in NIBs fits to embody an integrated access and backhaul network node for efficient spectrum utilization in a network.
- An NIB is targeting to build a network with high level of autonomy.

In the next section, we will present a case study on the network flexibility enhancing spectral efficiency in the mobile NIB-based maritime network.

IV. CASE STUDY: NETWORK FLEXIBILITY IN NIB-EMBODIED MARITIME NETWORK

The use cases of NIB are often considering the movement of end users in the network and the mobility of NIB is beneficial to exploit in such situation. We need to develop the optimal deployment of NIB for typical application scenarios of different use cases with respect to number (how many), location (where), time (when), and transport means (how) of NIBs. In line with this, we present in this section a case study to utilize the network flexibility of NIB in order to enhance the spectral efficiency of maritime communications by optimizing the path planning of a shipborne NIB in a NIB-based network, as shown in Fig. 3. The fleet of ships in the sea often suffer from weak network connectivity due to the lack of network infrastructure. NIB can improve this by diversifying the access into different networks and moving it to a favorable networking location.

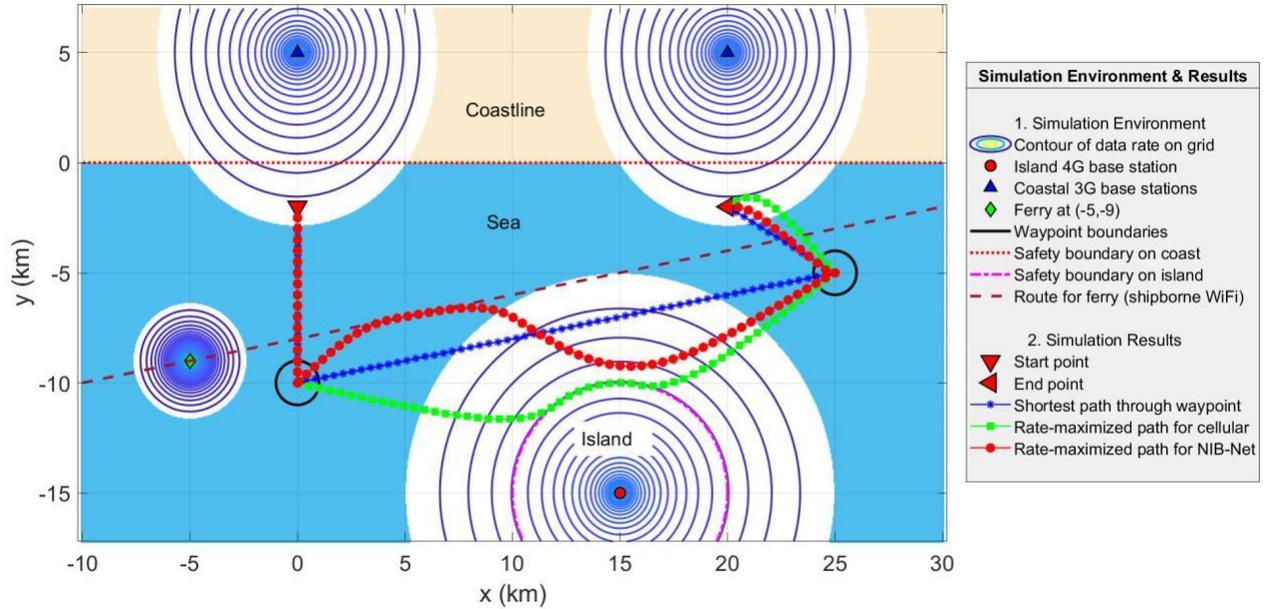

*Figure 3. Comparison of path planning for NIB-enabled ship fleet in maritime network. NIB-embodied ship fleet travel from start point to end point through two waypoints at maximum speed of 30 km/h. The ferry with shipborne WiFi passes on the route across the simulation region at the constant speed of 40 km/h.*

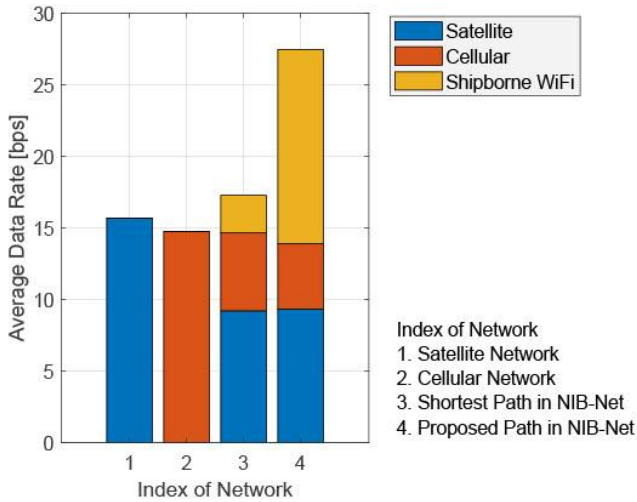

*Figure 4. Comparison of a portion of data rate contributed by each network for different maritime networks in Fig. 3; Network flexibility in NIB helps improve the spectral efficiency.*

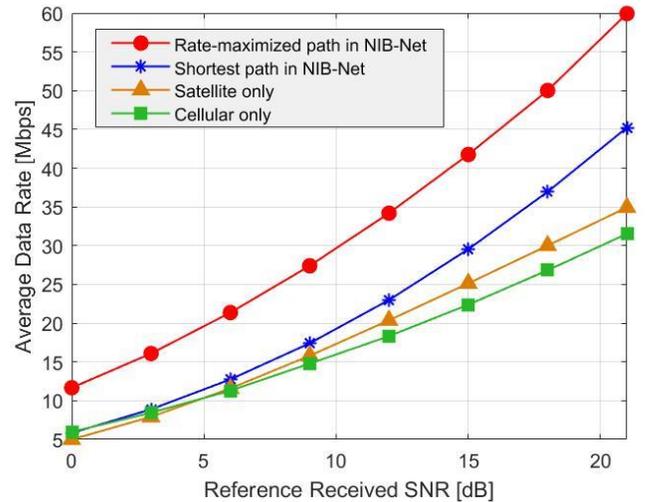

*Figure 5. Comparison of data rate with respect to the received SNR over different maritime networks; Network flexibility in NIB can facilitate the increased multiplexing gain at high SNR by taking advantage of larger bandwidth in ship-borne WIFI.*

For simulated networking environment, we assume that the NIB-enabled ship fleet have access to three maritime networks, i.e., satellite, coast-based and island-based cellular networks, and shipborne WiFi on ferry route. The system parameters are listed in Table I. For a total simulation time of two hours, the NIB-embodied ship fleet need to hover two waypoints spending 20 minutes, respectively, for marine activities, e.g., oceanographic sensing/monitoring, fishing, sightseeing, etc. We present three different optimized paths in Fig. 3; rate-maximized path and uniform speed shortest path in NIB-enabled network and rate-maximized path in cellular-only network for benchmark. Based on the network geometry and availability on time, the path planning of NIB-enabled ship fleet is dynamically varying to leverage network flexibility. Especially, with known information on ferry route, the NIB-enabled ship fleet can perform platooning with a ferry on the route for a while in order to connect to a high data rate shipborne WiFi. Consequently, the NIB-embodied maritime network can achieve significantly higher data rate than solely operating satellite and cellular networks, as shown in Fig. 4. The shortest path planning can significantly save the power consumption due to the stable movement of ship fleet, while still taking advantage of network flexibility against solely operational networks.

Fig. 5 illustrates the enhanced multiplexing gain when taking advantage of network flexibility leveraged by a NIB. NIB-embodied ship fleet help access to shipborne WiFi with higher bandwidth during a certain amount of simulation time. By doing so, it is clear that, as signal-to-noise ratio (SNR) increases, the slope of data rate in a NIB-enabled network is higher than that in a cellular-only or satellite-only network. NIB-leveraged ship fleet using both spectral efficient and power-efficient paths can benefit from the increased slope of rate due to the network flexibility.

## V. Challenges for Network-in-a-Box in 6G

In order to adopt NIB in 6G, the potential technologies stimulating 6G visions might be integrated with NIB. The transition toward 6G-NIB will face the technical challenges which are not considered in the traditional NIBs. 6G is constrained on more specified and service-centric requirements than 5G. NIBs should be evolved to fit the stringent requirements for new 6G services. For example, robotics and automated systems for industrial autonomy often require the extremely reliable connectivity with ultra-low delay for safety and precision issues. The lightweight and transportable NIB can play an important role to connect those entities in a private industrial sector.

6G services might be pursuing multi-purpose on not only communication but also control, computation, navigation, and sensing. This means that 6G-NIB should ideally deal with joint 3CLS service in a box independently and/or inter-operate with existing traditional network infrastructures including other NIBs as well as more heterogeneous multi-functional entities. Developing standard interfaces in a box connecting diverse multi-functional network components becomes a challenge.

The AI and ML technologies will be possible to adopt in NIB since they enable data-driven modeling which might be efficient in 6G network having massive dataset. For instance, the environmental parameters in the usage scenarios of NIB are hardly predictable and it is not cost-effective to estimate them in a traditional way since an NIB-embodied network is sometimes mobile and temporary. The learning-based network functionalities such as channel modeling and load balancing will help improve efficiency and quality-of-service in 6G-NIB networks.

Moreover, a few traditional NIBs were discussed to utilize the renewable energy sources since an NIB often works under the off-grid power condition. A 6G-NIB will face many circumstances where the energy sources are not sufficient or unavailable to guarantee 6G visions. The energy-efficient network functions in NIB should be developed and the energy generation profile should be integrated into NIB-based network protocol and architecture.

## VI. Conclusion

5G New Radio has been tested and being rolled out gradually all over the world and now we are in the initial phase for preparing new cellular generation era anticipating a real quantum leap forward for fully-digitalized society. 6G will connect everything into the global coverage ubiquitously and resiliently and alleviate the digital divide on the globe. To achieve this, the network will have to be organized, expanded, and, integrated in an on-demand fashion.

NIB is a disruptive solution most suitable to provide the connectivity-on-demand in the specific scenarios such as disaster recovery, intermittent events, challenging circumstances, etc. which might be overlooked for realizing ubiquitous connectivity in 6G. Network flexibility is the major feature for carrying out the NIB use cases, which can be characterized by easy deployment, technological diversity, and interoperability with external network entities. It can encourage NIBs to flexibly fulfill the 6G requirements. Meanwhile, 6G enabling technologies can strengthen the technological capability on NIB to effectively provision the on-demand services for different application scenarios.